\documentclass[english,aps,pop,reprint,superscriptaddress,showpacs,floatfix]{revtex4-1}
\usepackage{graphicx}
\usepackage{amsmath}
\usepackage{amssymb}
\usepackage{babel}
\usepackage{amsmath}
\usepackage{ulem}
\usepackage{pifont}
\usepackage{multirow}

\begin{document}

\begin{abstract}

Energy transport and confinement in tokamak fusion plasmas is usually determined by the coupled nonlinear interactions of small-scale drift turbulence and larger scale coherent nonlinear structures, such as zonal flows, together with free energy sources such as temperature gradients. Zero-dimensional models, designed to embody plausible physical narratives for these interactions, can help identify the origin of enhanced energy confinement and of transitions between confinement regimes. A prime zero-dimensional paradigm is predator-prey or Lotka-Volterra. Here we extend a successful three-variable (temperature gradient; microturbulence level; one class of coherent structure) model in this genre [M A Malkov and P H Diamond, Phys. Plasmas 16, 012504 (2009)], by adding a fourth variable representing a second class of coherent structure. This requires a fourth coupled nonlinear ordinary differential equation. We investigate the degree of invariance of the phenomenology generated by the model of Malkov and Diamond, given this additional physics. We study and compare the long-time behaviour of the three-equation and four-equation systems, their evolution towards the final state, and their attractive fixed points and limit cycles. We explore the sensitivity of paths to attractors. It is found that, for example, an attractive fixed point of the three-equation system can become a limit cycle of the four-equation system. Addressing these questions – which we together refer to as “robustness” for convenience, is particularly important for models which, as here, generate sharp transitions in the values of system variables which may replicate some key features of confinement transitions. Our results help establish the robustness of the zero-dimensional model approach to capturing observed confinement phenomenology in tokamak fusion plasmas.


\end{abstract}

\title{Robustness of predator-prey models for confinement regime transitions in fusion plasmas}

\author{H.~Zhu}
\affiliation{Centre for Fusion, Space and Astrophysics, Department of Physics, Warwick University, Coventry CV4 7AL, United Kingdom}

\author{S.C.~Chapman}
\affiliation{Centre for Fusion, Space and Astrophysics, Department of Physics, Warwick University, Coventry CV4 7AL, United Kingdom}

\author{R.O.~Dendy}
\affiliation{Euratom/CCFE Fusion Association, Culham Science Centre, Abingdon, Oxfordshire OX14 3DB, United Kingdom}

\maketitle

\begin{flushleft}
\textbf{1. Introduction}
\end{flushleft}

Energy transport in toroidal magnetically confined fusion plasmas is determined, in most cases, by the effects of small-scale turbulence and larger scale coherent nonlinear structures, together with their mutual interactions. These structures include zonal flows and geodesic acoustic modes\cite{HW87,CPB01,JFM02,M03,CSSRK05,DIIH05,GFMSS06}, which are radially localised poloidal flows, and streamers\cite{YIMK08}, which are radially highly elongated and poloidally localised. The importance of these structures for energy transport was highlighted in large scale numerical simulations\cite{DJKR00,JDKR00}, and the first direct experimental observation of streamers was reported in 2008\cite{YIMK08}. Zonal flows have been the subject of extensive theoretical and observational work\cite{HW87,CPB01,JFM02,M03,CSSRK05,DIIH05,GFMSS06}. There is now substantial experimental support for the long-standing hypothesis\cite{DK91} that the growth of zonal flows is driven by the averaged Reynolds stress of small scale turbulence. The latter can be locally suppressed by the resultant shear flow, thereby generating a temporally quasi-discontinuous enhancement  of global energy confinement: the L-H transition\cite{Wea82}. Whether zonal flows or streamers are preferentially formed under specific plasma conditions, and how they compete, has been addressed from various perspectives\cite{MRD01,LKMMD05,KYII08}, and remains an open question. For a recent review of experimental observations of the interaction between mesoscale structures (such as zonal flows and streamers) and microscale structures (such as drift turbulence), see\cite{F11}; of drift turbulence, particularly in relation to transitions in global confinement, see\cite{TFM09}; and of the L-H transition, see\cite{W07}.  A recent review of these physics issues in a broad context is provided by\cite{DHM11}. As emphasised in\cite{F11,TFM09,W07,DHM11} and references therein, recent diagnostic advances are transforming the experimental study of time evolving microturbulence and coherent nonlinear mesoscale structures during confinement transitions. This generates fresh theoretical challenges. In addition, the ability to understand and control this plasma physics phenomenology will be central to the successful operation of the next step magnetic confinement fusion experiment ITER\cite{Dea07}.

It is remarked by Malkov and Diamond in\cite{MD09}, hereafter referred to as MD, that transport models derived from the fundamental equations of plasma physics continue to add much to our understanding but ``tend to be increasingly, if not excessively, detailed.  Therefore, there is high demand for a simple, illustrative theoretical model with a minimal number of critical quantities responsible for the transition. Such models usually yield or encapsulate basic insight into complicated phenomena.'' One approach in fusion plasmas is that of zero-dimensional models for the interaction between microturbulence and coherent nonlinear structures, in particular predator-prey or Lotka-Volterra\cite{LCC93,DLCT94}. The properties of Lotka-Volterra systems, both mathematically and from the perspective of fusion plasma physics, are by no means fully explored and remain an active field of research\cite{HS94,BG03,SWA05,VWANS06,B10,II11}. For fusion applications, a key step is to establish agreement between the outputs of such models and the observed confinement phenomenology, which should ideally extend to the character of measured time traces of key properties near transitions, for example. Recent experimental results\cite{S12,XTD12} are encouraging in this respect. There is an important additional requirement. The zero-dimensional models used for this application should be robust, in the sense that the character of their outputs remains largely invariant against minor changes in the formulations of the models. This requirement for robustness has been explicitly noted\cite{WCDR99} in the other main class of zero-dimensional heuristic model for magnetised plasma confinement, namely sandpiles, both in fusion\cite{DH95,Nea96,Cea98,HelSOC98,CDR99,CDH01,GDR02} and in solar-terrestrial\cite{WCDR99,CWDHR98,HPDHM03,DCP07} contexts, and requires investigation for predator-prey and Lotka-Volterra applications to fusion plasmas.

There are several aspects to the degree of invariance of the phenomenology generated by a zero-dimensional model when aspects of the model are changed. First, what is the long-time behaviour of the system and how sensitive is this to variation in the model parameters\cite{H77,S84}? Second, how sensitively does the nature of the system's evolution towards its final state depend on the initial conditions? Is there an attractive fixed point or limit cycle towards which the system flows as time passes? If so, what is its basin of attraction? Third, how sensitive is the path to this attractor? This is particularly important for models which, as here, generate sharp transitions in the values of system variables which may replicate some key features of confinement transitions in tokamaks. If the initial conditions are varied, is the time at which the transition occurs delayed or brought forward, or does its character change, for example? Further, given two zero-dimensional models which are schematically distinct but adjacent, how similar is the phenomenology of their solutions? An example is provided here by our extension of the model of MD\cite{MD09} to incorporate two variables, rather than one, representing different classes of large scale coherent nonlinear field, in a four-variable system. The case of two predators and one prey was considered theoretically in the model of Itoh \& Itoh\cite{II11}, hereafter referred to as II, and by Miki and Diamond\cite{Miki2011}, and there is recent experimental motivation\cite{S12,XTD12}. Insofar as a zero-dimensional model turns out to be robust with respect to the considerations outlined (attractors; initial conditions; structural adjacency), confidence is strengthened in the mapping from model variables to specific plasma properties, and from the time evolving behaviour of the model to that of the plasma system.

In the present paper, we focus from this perspective on the interesting and successful mathematical model proposed in MD. This is constructed in terms of variables representing the magnitude of the plasma temperature gradient and the amplitudes of small scale drift turbulence and of large scale coherent nonlinear structures such as zonal flows. Malkov \& Diamond proposed\cite{MD09} certain mappings between different solution regimes of their model and different confinement regimes of tokamak plasmas. In the interest of continuity, we follow the confinement regime nomenclature of MD in relation to model outputs in the present paper. We investigate the robustness of the  phenomenology of the MD model extended as described, for parameter regimes identical, or adjacent, to those used in the key figures of MD. Where robustness is demonstrated and, if possible, explained, this reinforces confidence that models in the genre of MD and II may capture key features of the physics of confinement transitions in tokamak plasmas.

\begin{flushleft}
\textbf{2. Model equations}
\end{flushleft}

Specifically, the MD model is a closed system of nonlinear differential equations which couple the time evolution of three variables: the drift wave-driving temperature gradient \textit{N}, the energy density of drift wave turbulence \textit{E}, and the zonal flow velocity \textit{U}. The three variables of the II model exclude \textit{N}, retain drift turbulence energy density denoted by \textit{W}, and incorporate the energy densities of two competing classes of coherent nonlinear structure, zonal flows \textit{Z} and zonal fields (e.g. streamers) \textit{M}. Miki and Diamond\cite{Miki2011} introduced a zero-dimensional three-variable two-predator, one prey model, where the predators are identified with zonal flows and geodesic acoustic modes. The aspect of robustness which we first address can therefore be expressed in physical terms as follows. We adopt the philosophy of II and of Ref.\cite{Miki2011} by introducing two competing classes of coherent nonlinear structure, here identified with zonal flows and streamers, that replace the single class in MD. The other two MD equations are adjusted only so far as necessary to accommodate these two fields, instead of one, in a mathematically symmetrical way as in II. We investigate how far the model outputs of our new four-variable system differ from those of the three-variable system of MD. A good focus for this study is provided by the time traces captured in Figs.2-4 of MD, which have been mapped to transitions observed between tokamak confinement regimes. How are these traces altered by the inclusion of a second competing class of coherent nonlinear structure? The answers to these questions are conditioned by the underlying phase space structure of families of solutions to the models, as plotted in Fig.5 of MD, for example. In addition to studying time traces, therefore, we seek to characterise the limit cycles and fixed points of our system of equations. We first generalize the un-normalized MD equations to:

\begin{eqnarray}
\frac{d\mathcal{E}}{d\tau} &=&\left(\mathcal{N} - a_{1}\mathcal{E} - a_{2}d^{2}\mathcal{N}^{4} - a_{3}V_{ZF}^{2} - a_{3}V_{SF}^{2}\right) \mathcal{E}\\
\frac{dV_{ZF}}{d\tau} & = & \left(\dfrac{b_{1Z}\mathcal{E}}{1 + b_{2Z}d^{2}\mathcal{N}^{4}} - b_{3Z}\right)V_{ZF} \\
\frac{dV_{SF}}{d\tau} & = & \left(\dfrac{b_{1S}\mathcal{E}}{1 + b_{2S}d^{2}\mathcal{N}^{4}} - b_{3S}\right)V_{SF} \\
\frac{d\mathcal{N}}{d\tau} &=& - \left(c_{1}\mathcal{E} + c_{2}\right)\mathcal{N} + q(\tau)
\end{eqnarray} 

This model encompasses drift wave turbulence level $\mathcal{E}$, drift wave driving temperature gradient $\mathcal{N}$, zonal flow velocity $V_{ZF}$, streamer flow velocity $V_{SF}$, and the heating rate $q$ which is a control parameter of the system. This model thus extends, to the case when zonal flows are joined by streamers, the key physics encapsulated in the description in\cite{KD03}: ``When the drift wave turbulence drive becomes sufficiently strong to overcome flow damping, it generates zonal flows by Reynolds stress. Drift wave turbulence and zonal flows then form a self-regulating system as the shearing by zonal flows damps the drift wave 
turbulence.'' We note that this model follows the approach expressed in Eq.(17) of MD\cite{MD09}, in that the zonal flows and streamers do not explicitly enter the time evolution equation for the temperature gradient, Eq.(4). The zonal flows and streamers are indirectly coupled to each other through the evolving temperature gradient and microturbulence level. To maximise mathematical congruence with the model of MD, there is no direct cross term in $V_{SF}V_{ZF}$.

The corresponding normalized equations are

\begin{eqnarray}
\frac{dE}{dt} &=& \left(N - N^4 - E - U_{1} - U_{2}\right) E\\
\frac{dU_{1}}{dt} & = & \nu_{1}\left( \dfrac{E}{1 + \zeta N^4} - \eta_{1}\right)U_{1} \\
\frac{dU_{2}}{dt} & =& \nu_{2}\left( \dfrac{E}{1 + \zeta N^4} - \eta_{2}\right)U_{2}\\
\frac{dN}{dt} &=& q - \left( \rho + \sigma E\right)N
\end{eqnarray} 
Here we have defined normalized variables

\begin{center}
$N=a_{2}^{1/3}\mathcal{N}$, $E=a_{1}a_{2}^{1/3}\mathcal{E}$, $U_{1}=a_{2}^{1/3}a_{3}V_{ZF}^{2}$, $U_{2}=a_{2}^{1/3}a_{3}V_{SF}^{2}$,  $t=a_{2}^{-1/3}\tau$,
\end{center}
and the transformed model parameters are

\begin{center}
$\nu_{1}=\dfrac{2b_{1Z}}{a_{1}}$, $\nu_{2}=\dfrac{2b_{1S}}{a_{1}}$, $\eta_{1}=\dfrac{b_{3Z}}{b_{1Z}}a_{1}a_{2}^{1/3}$, $\eta=\dfrac{b_{3S}}{b_{1S}}a_{1}a_{2}^{1/3}$, $\rho=c_{2}a_{2}^{1/3}$, $\sigma=\dfrac{c_{1}}{a_{1}}$, $\zeta=\dfrac{b_{2}}{a_{2}^{4/3}}$, $q(t)=a_{2}^{2/3}q(\tau)$, $d=1$.
\end{center}
We note that this is equivalent to the normalization of MD only when $a_{2}= 1$. This reflects a minor inconsistency in the normalization choice of MD. 
The system of Eqs.(5-8) thus generalizes the system of Eqs.(15-17) of MD by introducing two distinct flow variables, $U_{1}$ and $U_{2}$, to replace the single zonal flow variable $U$. We refer to $U_{1}$ as zonal flow, $U_{2}$ as streamer flow.

Section 3 of this paper addresses transition phenomenology given time-independent coefficients, as characterised primarily by time traces. This requires careful comparison with the specific scenarios identified in Fig.3 to Fig.5 of MD. The MD scenarios predetermine the choice of parameter values and initial conditions that we consider. We typically probe neighbouring phase space by considering in addition eighty-one (three to the fourth power) nearby phase trajectories. In Section 4 we consider the phase space evolution of our system and establish comparisons between the MD model and ours. In Section 5 we analyse possible links to the phenomenology of tokamak plasmas, in the spirit of MD and II.

\mbox{}\\
\textbf{3. Modelling confinement transitions}
\mbox{}\\

In the limit where either one of the two parameters that represent distinct classes of coherent nonlinear structures (zonal flows or streamers) in our model vanishes, it reproduces exactly the results shown in Fig.2 of MD, as required. Figure 1 displays the corresponding results for the case where both streamers and zonal flows exist. In the nomenclature of MD, the system starts from an overpowered state near H-mode, with negligible turbulence $E$ and large scale structures $U_{1}$, $U_{2}$. The eventual growth of turbulence accompanies a sharp drop in $N$ to unstable L-mode, while also providing energy for $U_{1}$ and $U_{2}$. Drift wave turbulence is later suppressed and the maximum amplitude of large scale flows declines, leaving only the mean flow to support the transport barrier\cite{DHM11}. Finally the stable T-mode, which combines a steady-state level of $E$ with lower $N$ than H-mode, appears after the oscillating transition regime. During this transition, energy is extracted from the initially dominant oscillating streamer flow $U_{2}$ to the zonal flow $U_{1}$ until the former vanishes.

In Fig.2, we plot the system evolution for the case where the values of $\nu_{2}$ and $\eta_{2}$ are different from Fig.1, while all other parameter values are identical. Specifically, in Fig.1 $\nu_{2}/\nu_{1} = \eta_{2}/\eta_{1} = 1.01$, whereas in Fig.2 $\nu_{2}/\nu_{1} = 0.01$ and $\eta_{2}/\eta_{1} = 0.1$. This weakens both the drive and the damping of structures $U_{2}$ compared to zonal flows $U_{1}$ in Fig.2, with respect to the case of Fig.1. Before time reaches t $\sim$ 6000, the evolution is very similar to Fig.2 of  MD. However, at t$\sim$ 6500 we find a dramatic change. A limit cycle appears after the long-term fixed point time series. The amplitudes of $U_{1}$ and $U_{2}$ exchange rather fast compared to Fig.1. Furthermore, the period of the limit cycle is rather long: several hundred time units. With the appearance of zonal flows and streamers, the T-mode becomes unstable.

Figure 3 shows the case where the heating rate is higher than for Fig.1, $q = 0.58$, but all other model parameters are the same. At each pulsed occurrence of zonal flows $U_{1}$ and streamers $U_{2}$, the former extract energy from the latter, which become extinct after the sixth pulse. Thereafter there are limit cycle oscillations in $E$, $N$ and $U_{1}$ equivalent to the limit cycle for $E$, $N$ and $U$ in the case in MD.

Figure 4 shows time traces for the case where all parameters, except the heating rate $q = 0.58$ which is the same as in Fig.3, are those of Fig.2. Together with Fig.5, where the heating rate $q$ is slightly increased to $q = 0.582$ instead of $q = 0.58$, this enables us to relate our model to Fig.4 of MD, which showed that if in MD $q = 0.582$ instead of 0.58, the limit cycle eventually collapses after many oscillations. The final state has $N$ finite and the remaining variables are zero; this is designated the QH-mode fixed point in MD. The corresponding cases for our model Eqs.(5-8) are shown in Figs.4 and 5. A precursor to limit cycle collapse is apparent in Fig.4 in the growth of the streamer field $U_{2}$ during the episodes of zonal flow quiescence in the last few oscillations of the system.

For the slightly different parameter set used to generate Fig.5, the pulses of $U_{1}$ and $U_{2}$ grow and die together. Their peak amplitude increases at each successive cycle, as does the time interval between them. At the final oscillation, $U_{1}$ and $U_{2}$ collapse promptly together, whereas E survives longer until it is extinguished by damping. The phenomenology of Fig.5 thus corresponds more closely to that of Fig.4 of MD, compared to our Fig.4.

Figure 6 illustrates how system evolution towards the finite-$N$ final state of Fig.5 depends on the damping rate $\eta_{2}$ of streamers. We fix all parameters except $\eta_{2}$ and find that, with increasing $\eta_{2}$, there are more peaks of $U_{2}$ correlating with cyclic growth of $E$, which acts as a damping sink of $N$. Successive peaks increase in height prior to extinction, which results in a final state similar to Fig.5.

\begin{figure}
\begin{center}
\includegraphics[width=1\columnwidth]{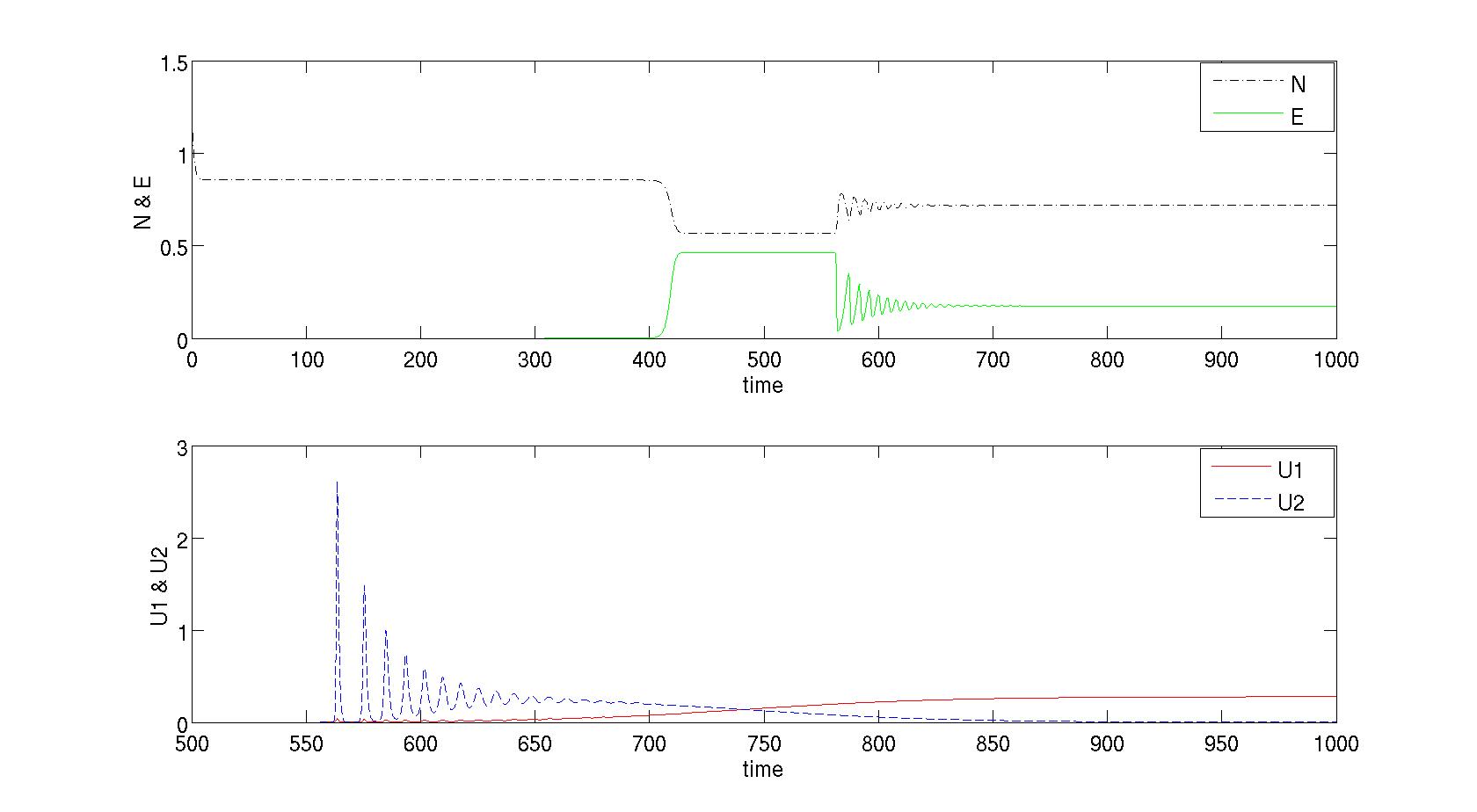}
\caption{Upper panel: From overpowered H-mode to unstable L-mode then to T-mode. Lower panel: Transition to T-mode for $U_{1}$ and $U_{2}$ showing intersection at $t\simeq 750$ followed by energy reversal. The parameters are $\nu_{1}=19$, $\nu_{2}=1.01\nu_{1}$, $\eta_{1}=0.12$, $\eta_{2}=1.01\eta_{1}$, $q=0.47$, $\rho=0.55$, $\sigma=0.6$, $\zeta=1.7$.}
\end{center}
\end{figure}

\begin{figure}
\begin{center}
\includegraphics[width=1\columnwidth]{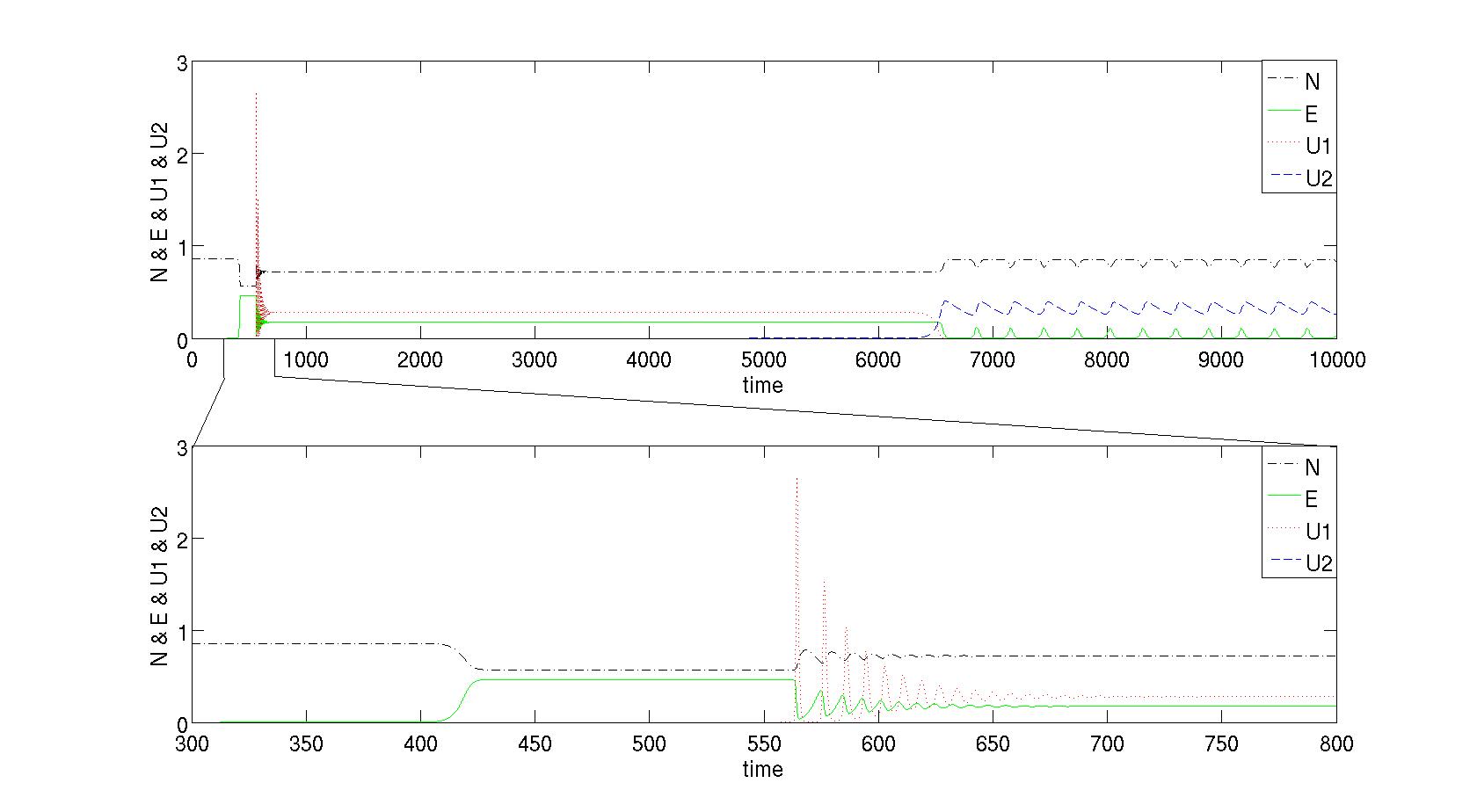}
\caption{Upper panel: Transition from stable fixed point state to unstable oscillatory limit cycle state. Lower panel: Zoom in version from $t=300$ to $t=800$. The parameters are $\nu_{1}=19$, $\nu_{2}=0.01\nu_{1}$, $\eta_{1}=0.12$, $\eta_{2}=0.1\eta_{1}$, $q=0.47$, $\rho=0.55$, $\sigma=0.6$, $\zeta=1.7$.}
\end{center}
\end{figure}

\begin{figure}
\begin{center}
\includegraphics[width=1\columnwidth]{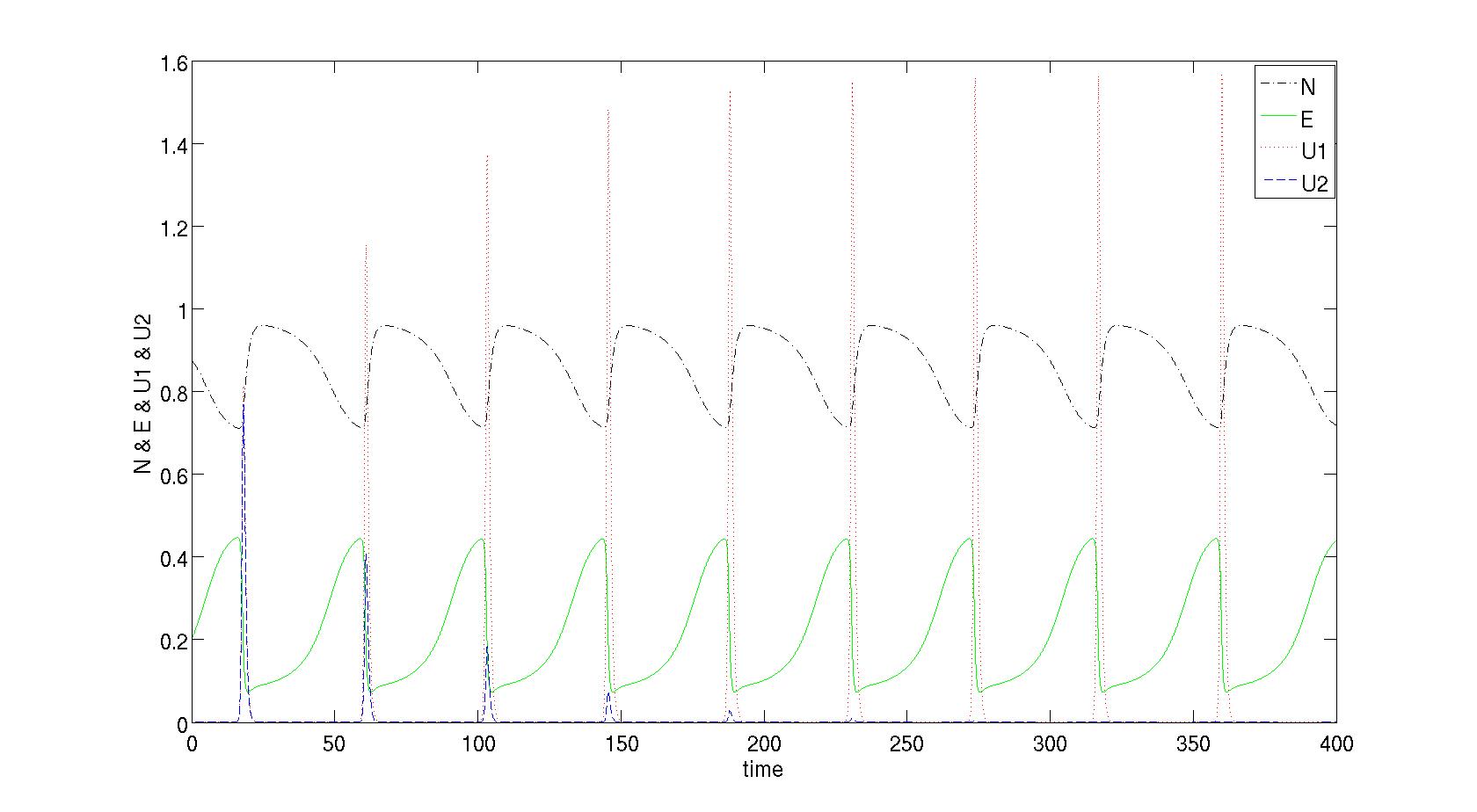}
\caption{Energy transfer from $U_{2}$ to $U_{1}$ during pulses of strong nonlinear oscillation, followed by limit cycle ocillation in $N$, $E$ and $U_{1}$. The parameters are $\nu_{1}=19$, $\nu_{2}=1.01\nu_{1}$, $\eta_{1}=0.12$, $\eta_{2}=1.01\eta_{1}$, $q=0.58$, $\rho=0.55$, $\sigma=0.6$, $\zeta=1.7$.}
\end{center}
\end{figure}

\begin{figure}
\begin{center}
\includegraphics[width=1\columnwidth]{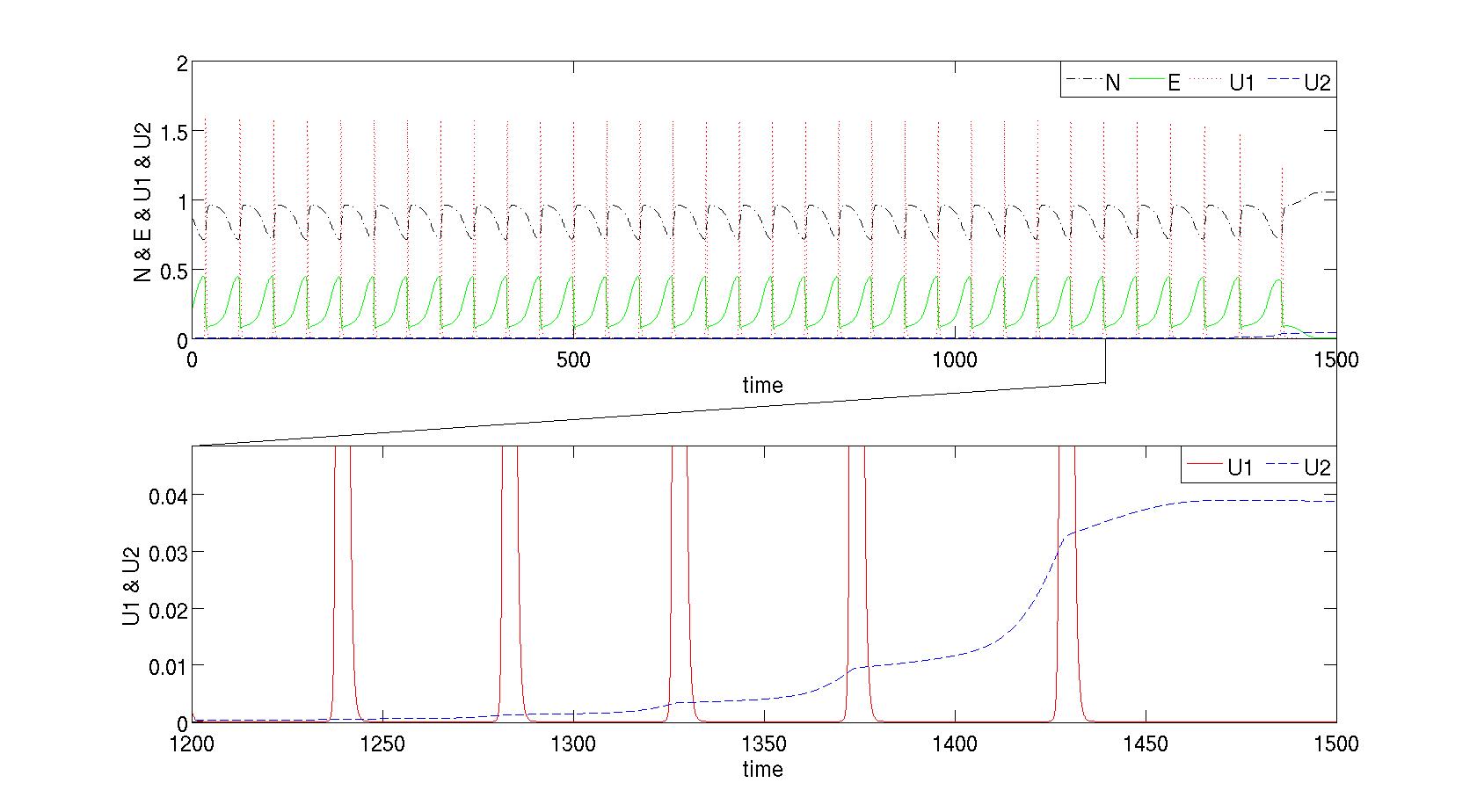}
\caption{Upper panel: Collapse of limit cycle in $N$, $E$ and $U_{1}$. Lower panel: Stair increasing of $U_{2}$. The parameters are $\nu_{1}=19$, $\nu_{2}=0.01\nu_{1}$, $\eta_{1}=0.12$, $\eta_{2}=0.01\eta_{1}$, $q=0.58$, $\rho=0.55$, $\sigma=0.6$, $\zeta=1.7$.}
\end{center}
\end{figure}

\begin{figure}
\begin{center}
\includegraphics[width=1\columnwidth]{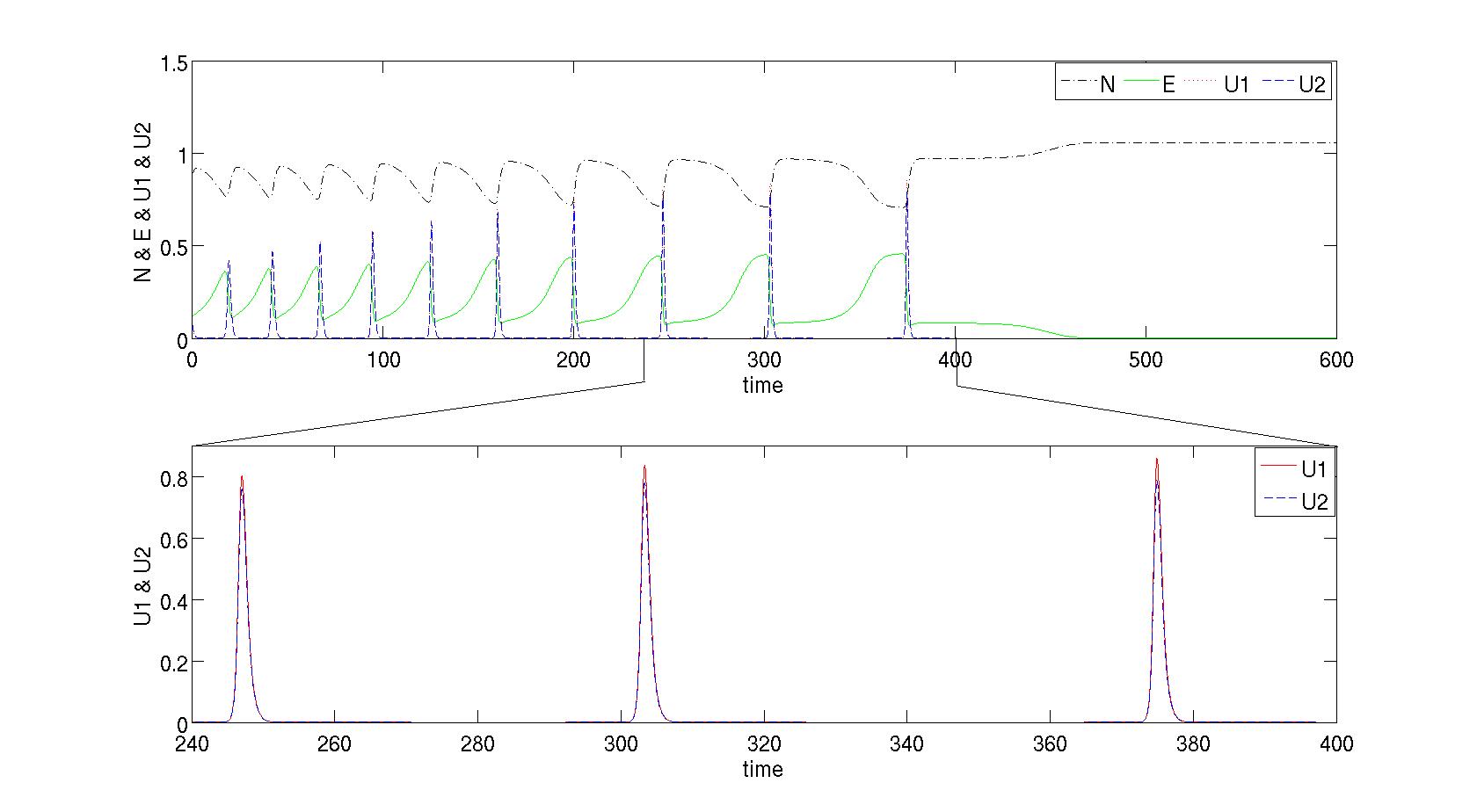}
\caption{Upper panel: Collapse of limit cycle with positively correlated growth of pulses of $U_{1}$ and $U_{2}$. Lower panel: Zoom in version from $t=240$ to $t=400$. The parameters are $\nu_{1}=19$, $\nu_{2}=1.0001\nu_{1}$, $\eta_{1}=0.12$, $\eta_{2}=1.0001\eta_{1}$, $q=0.582$, $\rho=0.55$, $\sigma=0.6$, $\zeta=1.7$.}
\end{center}
\end{figure}

\begin{figure}
\begin{center}
\includegraphics[width=1\columnwidth]{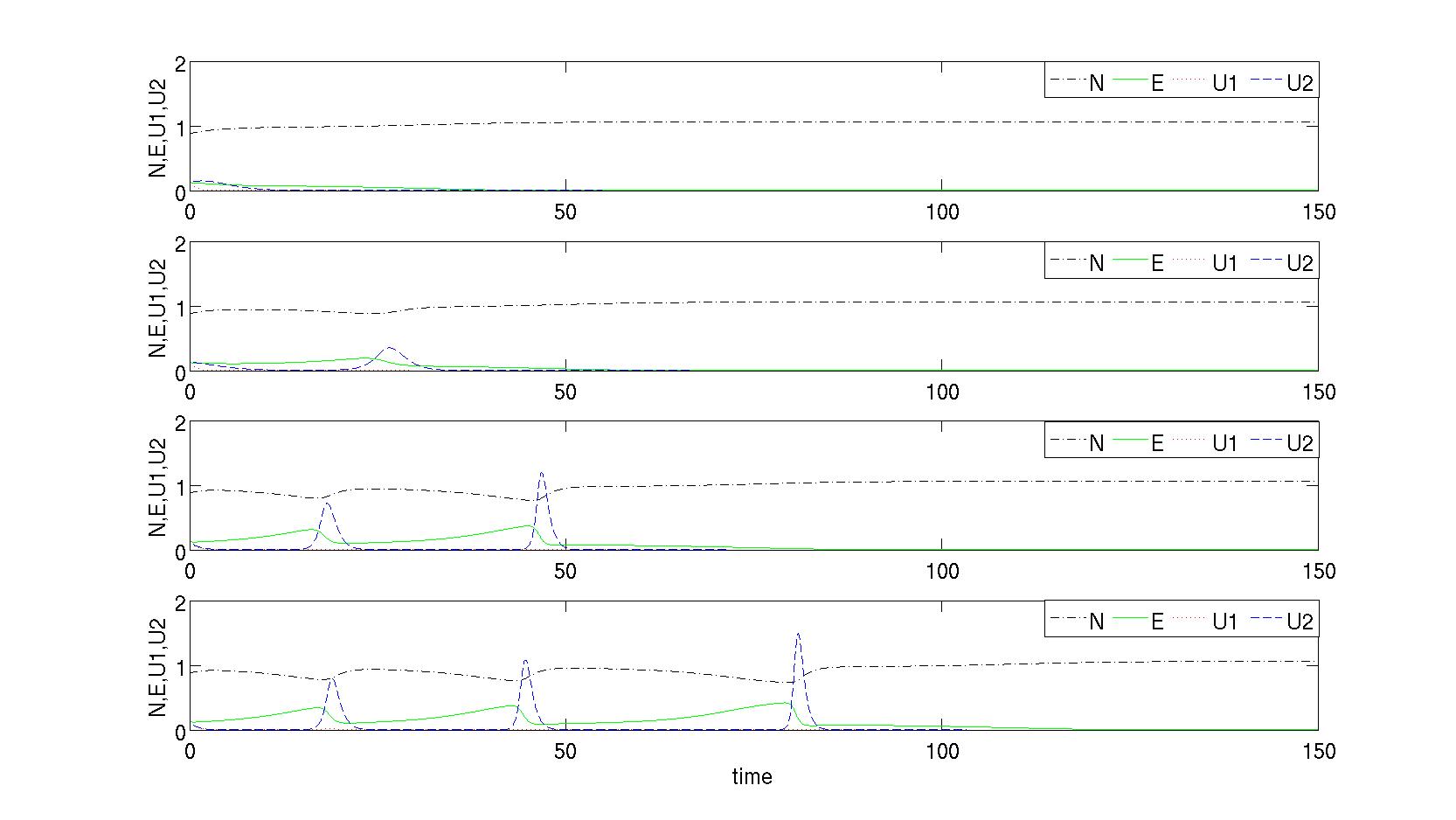}
\caption{Evolution to the finite $N$ attractor for different values of $\eta_{2}$. Upper panel: $\eta_{2}=0.05$. Middle upper panel: $\eta_{2}=0.06$. Middle lower panel: $\eta_{2}=0.10$. Lower panel: $\eta_{2}=0.11$. The remaining parameters are the same: $\nu_{1}=19$, $\nu_{2}=1.001\nu_{1}$, $\eta_{1}=0.12$, $q=0.582$, $\rho=0.55$, $\sigma=0.6$, $\zeta=1.7$.}
\end{center}
\end{figure}

\mbox{}\\
\textbf{4. Phase space evolution}
\mbox{}\\

The time traces of the individual variables, plotted in Figs.1 to 6, represent projections of the evolution in four-dimensional phase space of the system defined by Eqs.(5) to (8). In the present section, we capture the global phase space explored by this system, for parameter values corresponding, or adjacent, to those used to generate Figs.1 to 6. This approach enables us to identify and characterise the nature of initial and final states, and of the transitional behaviour between them. These results are supplemented in the Appendix by stability studies. At issue are two main physical concerns, which map directly to the properties of different energy confinement regimes in tokamaks, insofar as the zero-dimensional approach and the identifications made in MD, for example, may be valid. First, what is the nature of the final state that is reached at long times? For example, is it an attractive fixed point or a limit cycle (implying a nearby repulsive fixed point)? Second, there is the question, discussed previously, of robustness of three-variable models against the inclusion of a fourth variable (here, streamers) in the model. For example, the pioneering work of MD includes identification of a limit cycle (Fig.3 of MD) with a specific confinement regime. Is this limit cycle - and, proceeding by analogy, the confinement regime that it represents - stable against the presence of streamers in addition to zonal flows?

Figure 7 displays the generalisation, to the four-variable system, of the case of the three-variable system addressed in Fig.2 of MD. To fix ideas, the two left-hand plots correspond to the three-variable case for the parameters of Fig.2 of MD, showing the attractive fixed point which has finite values of $E$, $N$ and $U$. The inward spiral path of the system from a random initial position is shown, both in ($E$, $N$, $U$) space and projected onto the ($E$, $U$) plane. It is evident that this path lies on a topological structure in phase space, whose dimensionality is lower by one than that of the full phase space. The two right-hand plots of Fig.7 show how this system changes when the two variables $U_{1}$ and $U_{2}$ replace $U$, for the parameter values used to generate the traces in Fig.1, which are adjacent to those for Fig.2 of MD, as discussed above. The centre right-hand plot shows initial spiral convergence in ($E$, $U_{2}$) which closely resembles that in the ($E$, $U$) plane displayed at centre left. Whereas with three variables this convergence is towards a fixed point, the existence of a fourth variable renders this attractive fixed point unstable. In consequence, the final stage of system evolution consists of injection in the $U_{1}$ direction to a fixed point at finite ($E$, $N$, $U_{1}$) with $U_{2} = 0$. The far right plot in Fig.7 demonstrates that this is indeed a fixed point, towards which phase space evolution originating from eighty-one different initial points converges. In each case, there is spiral convergence on a manifold followed by injection along $U_{1}$. The choice of initial condition affects only the orientation of this convergence manifold with respect to $U_{1}$ and $U_{2}$. We note also that the final state with finite $U_{1}$ differs from the MD final state for which $U = 0$.

Figure 8 illustrates the phase space evolution of the system whose time traces are plotted in Fig.2, which like Fig.7 is a case with parameters adjacent to those used to generate Fig.2 of MD. The initial spiral convergence in the ($E$, $U_{1}$) plane, shown in the centre panel, resembles that in the ($E$, $U$) plane for the MD case plotted in the left panel, which is identical to the centre-left panel of Fig.7. As in Fig.7, the stable fixed point of the three-variable system is unstable for the four-variable system, for which there is injection along $U_{2}$. Unlike Fig.7, where this injection is towards a stable fixed point, in Fig.8 the injection is onto a stable limit cycle that has finite slow oscillations in ($N$, $E$, $U_{2}$) with $U_{1} = 0$ in the four-variable system.

The three-variable MD system has a limit cycle in ($N$, $E$, $U$) for the case shown in Fig.3 of MD. This is re-plotted in the two left panels of Fig.9 and in the left panel of Fig.10. Figures 9 and 10 relate to the time traces shown in Figs.3 to 5 of the present paper, obtained for parameter sets for the four-variable system which are adjacent to those used in MD for the three-variable system. For the parameters of Fig.9, which is the phase space plot for Fig.3, it is clear from the two right-hand panels that the limit cycle behaviour is essentially that of the MD system. The transient evolution towards the limit cycle involves circulation on similar planes that have successively lower peak values of $U_{2}$. The final limit cycle in ($N$, $E$, $U_{1}$), with $U_{2} = 0$, is essentially that in ($N$, $E$, $U$) for the three-variable system. 

The three-variable MD attractive limit cycle which manifests in the four-variable system as shown in Fig.9 is, however, unstable. Figure 10, which is the phase space plot for Fig.4, shows that the system leaves the former limit cycle and transiently explores the additional phase space dimension associated with the additional variable, before converging to a new fixed point that has $N$ finite and all other variables zero. This class of attractive fixed point is noted in Fig.4 of MD, shown in the far left panel of Fig.11 and, projected on the ($E$, $U$) plane, in the centre left panel. The two right-hand panels of Fig.11 are the phase space plots for Fig.5, showing convergence to the origin in ($E$, $U_{1}$, $U_{2}$) space while $N$ remains finite. The final step to the origin is preceded by circulation around and away from an apparent repulsive fixed point with finite values of $E$, $U_{1}$ and $U_{2}$. The far right panel of Fig.11 shows that the choice of initial conditions merely affects the orientation in ($U_{1}$, $U_{2}$) space of the plane of this transient circulation.

The phase space behaviour discussed thus far assists us in re-visiting the time traces in Fig.2, for which the corresponding phase plot is given in Fig.13. In Fig.12 we annotate Fig.2 in light of Fig.13. These two Figures demonstrate how, for the four-variable system, the T-mode of the three-variable system becomes unstable at long times. The system then evolves towards the newly identified attractive limit cycle in ($N$, $E$, $U_{2}$). Here slow oscillations in $N$ correlate with those in $U_{2}$, both of which remain finite throughout, while bursts of $E$, feeding on $U_{2}$, occur between extinctions.

\mbox{}\\
\textbf{5. Conclusions}
\mbox{}\\

Contemporary experimental results from the DIII-D\cite{S12} and HL-2A tokamaks\cite{XTD12} reinforce the relevance of zero-dimensional predator-prey models to transitions between energy confinement regimes. Understanding how the outputs of related, but different, predator-prey models for plasma confinement phenomenology may resemble or deviate from each other is therefore important. In this paper we have focused on the consequences of adding a second predator, and hence a fourth field variable, to the three-field MD\cite{MD09} model. Quantitative studies have been presented for parameter sets that are maximally adjacent to those in MD, which yield the time traces shown in Figs.1 to 6 and Fig.12. These are projections of the phase space dynamics shown in Figs.7 to 11 and Fig.13. It is found that both congruences and deviations can occur between the three-field and four-field models. For example, Fig.10 shows how a limit cycle in the three-field system is unstable for four fields in the relevant parameter range, where the attractor is a fixed point. Conversely Fig.8 shows a three-field fixed point mapping to a four-field limit cycle. Figure 13 shows the complex, but resolved, phase space dynamics underlying a generalisation to four fields of the three-field scenario modelled in Fig.2 of MD. We conclude that exploration of the linkages between different zero-dimensional models, capturing full phase space properties so far as computationally possible, needs to keep pace with the continuing development and refinement of individual zero-dimensional models in fusion plasma physics.

Zero-dimensional models remain attractive because they embody physically motivated narratives that may account for global fusion plasma confinement phenomenology. Ideally the end states (attractors) of zero-dimensional models, together with the transitional behaviour en route from the initial configurations, should be robustly identifiable with fusion plasma confinement states and transitions. Zero-dimensional predator-prey models, constructed in terms of a small number of variables representing global quantities such as the drift wave turbulence level $\mathcal{E}$, drift wave driving temperature gradient $\mathcal{N}$, zonal flow velocity $V_{ZF}$, streamer flow velocity $V_{SF}$, and the heating rate $q$ in Eqs.(1) to (4), are intrinsically nonlinear. This nonlinearity implies the potential for a rich and varied set of attractors and transitional behaviour, together with strong dependence on the numerical values of model parameters. The present paper has taken steps to explore this potential for the model of interest in the case of parameter sets close to those studied previously in MD, with a view to strengthening the links between families of zero-dimensional models on the one hand, and fusion plasma confinement phenomenology on the other. We note finally that some of the considerations addressed here may carry over to other fields where it is hoped to develop zero-dimensional models that have descriptive, or even predictive, power for global phenomena in macroscopic multiscale driven-dissipative systems. A topical instance is provided by zero-dimensional modelling in climate science, see for example Ref.\cite{EliseevClimate2006} and references therein, where some general circulation models incorporate Lotka-Volterra features\cite{CoxClimate2000}.

This work was part-funded by the EPSRC and the RCUK Energy Programme under grant EP/I501045 and the European
Communities under the contract of Association between EURATOM and CCFE. The views and opinions expressed
herein do not necessarily reflect those of the European Commission.

\newpage
\begin{figure*}
\begin{center}
\includegraphics[width=2\columnwidth]{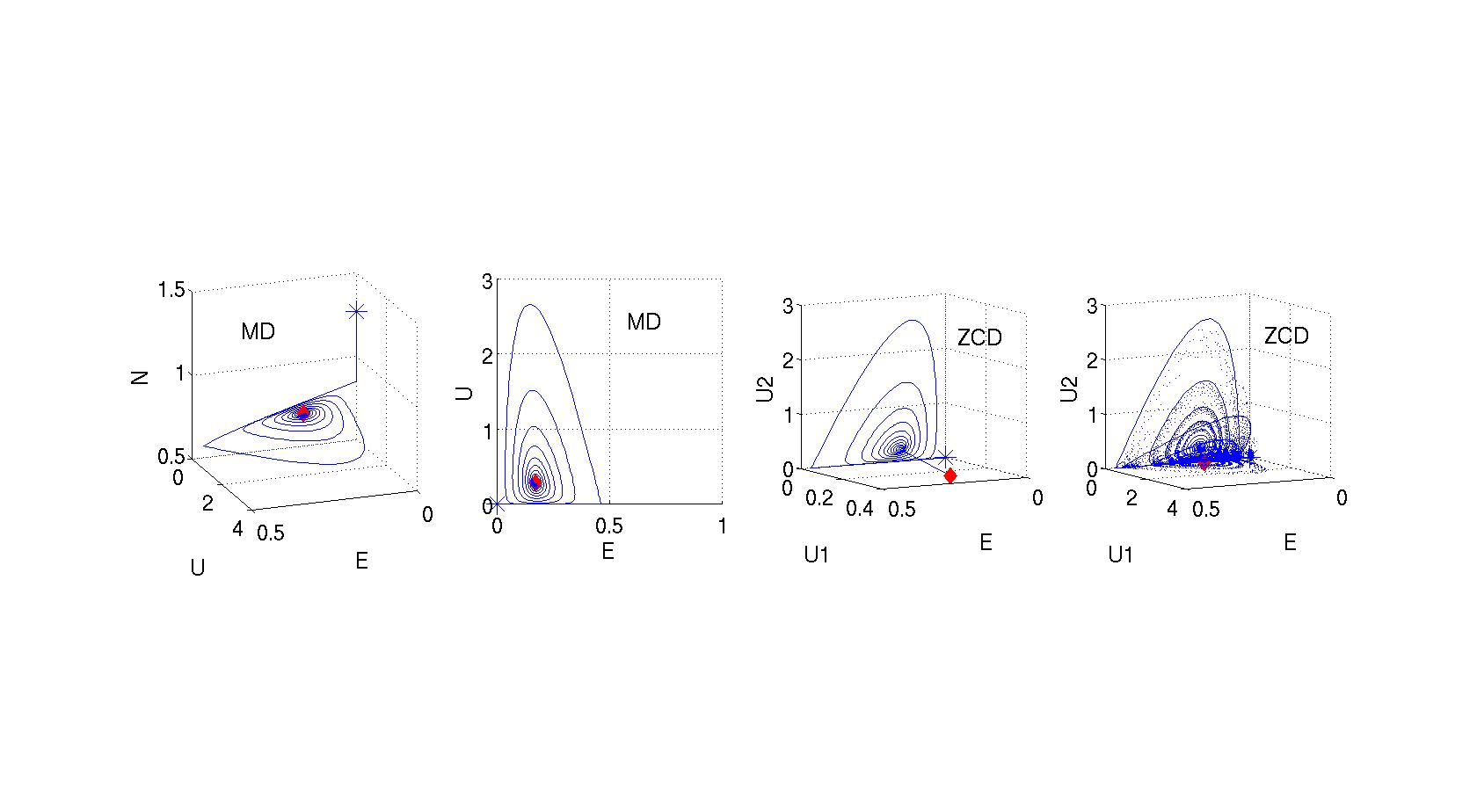}
\caption{First panel: Fig.2 in MD. The parameters are $\nu=19$, $\eta=0.12$, $q=0.47$, $\rho=0.55$, $\sigma=0.6$, $\zeta=1.7$. Second panel: Projection of first panel on $E$-$U$ plane. Third panel: Phase plot of Fig.1. Last panel: Phase plot of Fig.1 with 81 initial conditions. Stars denote initial values, blue dots denote trajectories and red diamonds denote final states.}
\end{center}
\end{figure*}

\begin{figure*}
\begin{center}
\includegraphics[width=2\columnwidth]{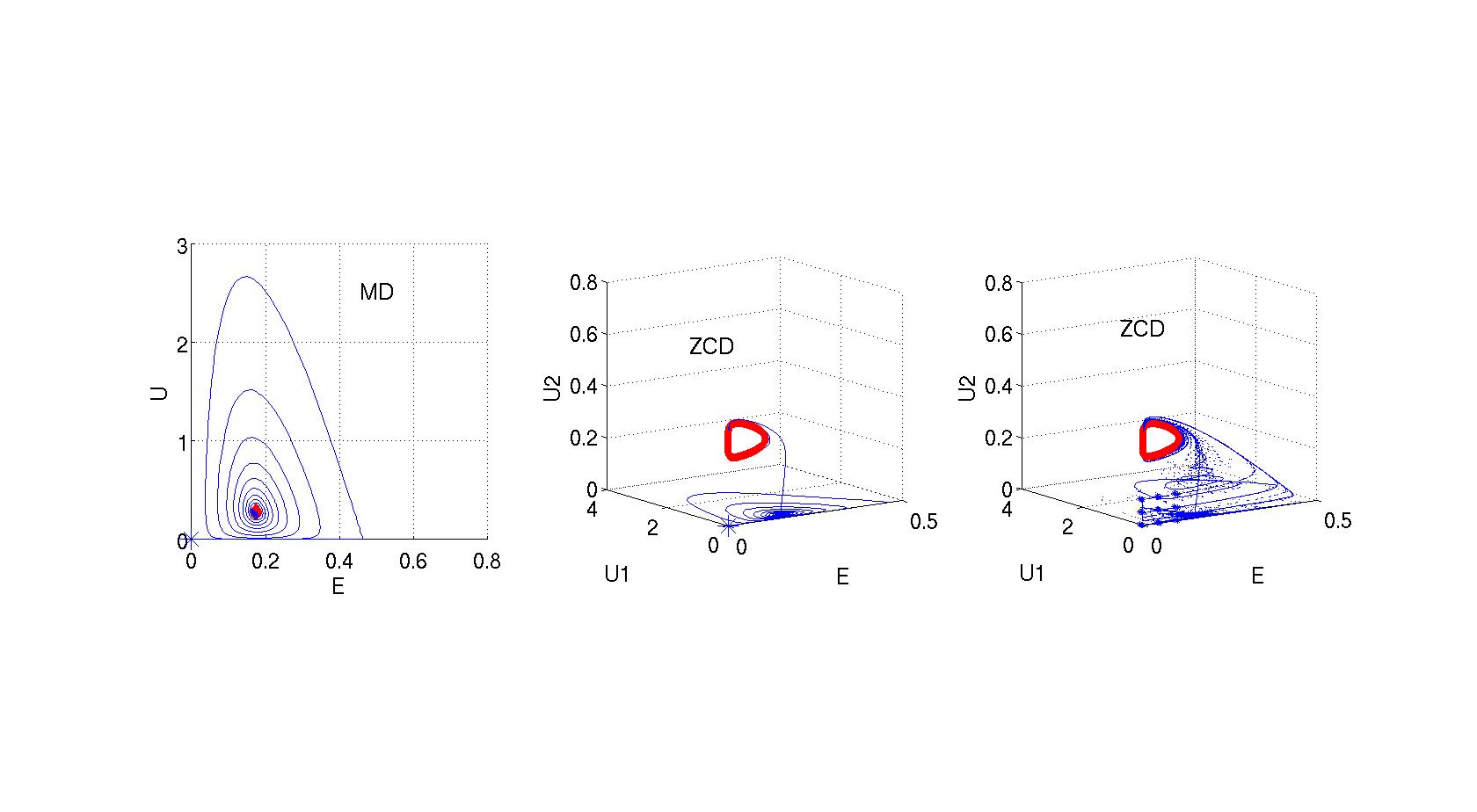}
\caption{First panel: Projection of Fig.2 in MD on $E$-$U$ plane. The parameters are $\nu=19$, $\eta=0.12$, $q=0.47$, $\rho=0.55$, $\sigma=0.6$, $\zeta=1.7$. Second panel: Phase plot of Fig.2. Last panel: Phase plot of Fig.2 with 81 initial conditions. Stars denote initial values, blue dots denote trajectories and red diamonds denote final states.}
\end{center}
\end{figure*}

\begin{figure*}
\begin{center}
\includegraphics[width=2\columnwidth]{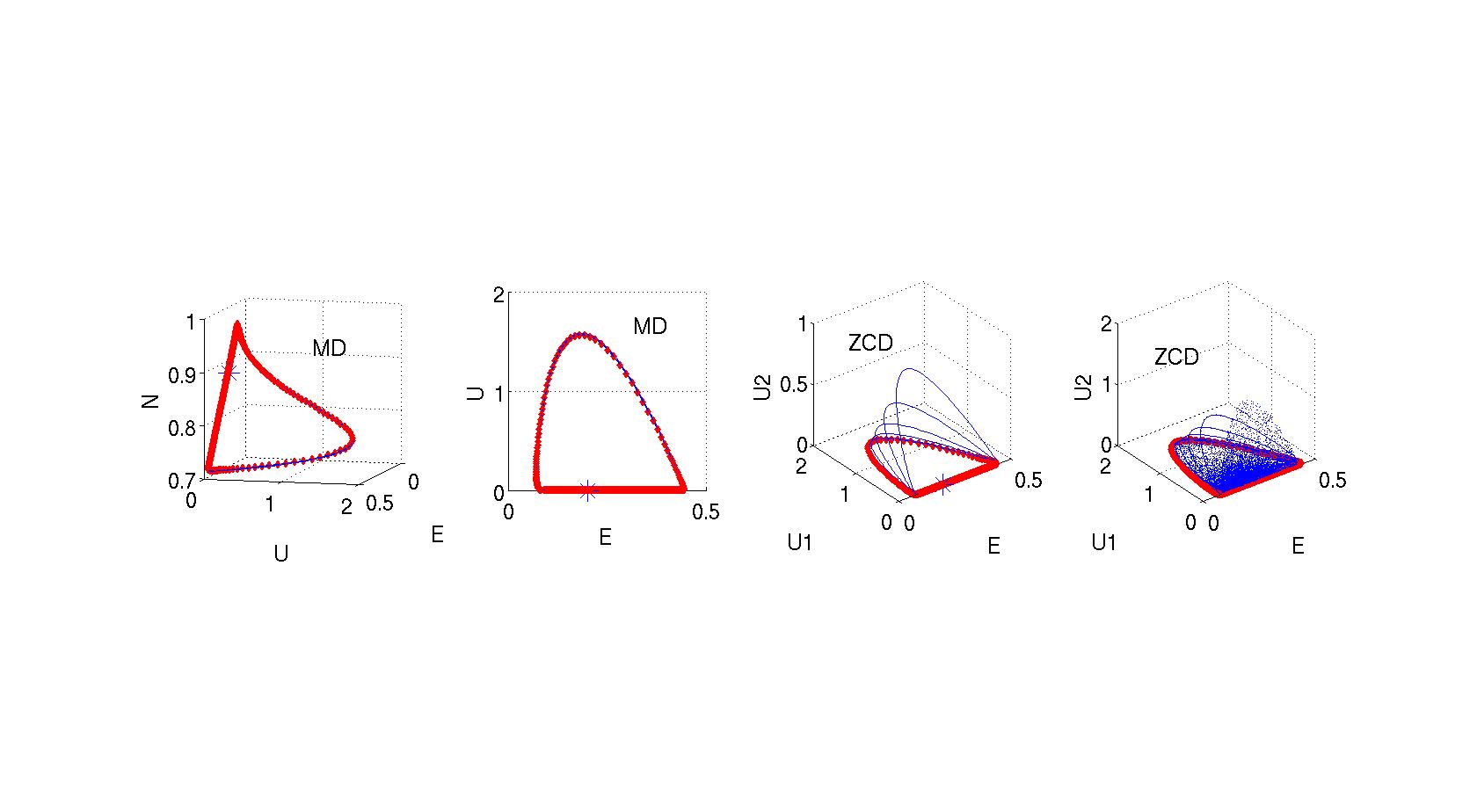}
\caption{First panel: Fig.3 in MD. The parameters are $\nu=19$, $\eta=0.12$, $q=0.58$, $\rho=0.55$, $\sigma=0.6$, $\zeta=1.7$. Second panel: Projection of first panel on $E$-$U$ plane. Third panel: Phase plot of Fig.3. Last panel: Phase plot of Fig.3 with 81 initial conditions. Stars denote initial values, blue dots denote trajectories and red diamonds denote final states.}
\end{center}
\end{figure*}

\begin{figure*}
\begin{center}
\includegraphics[width=2\columnwidth]{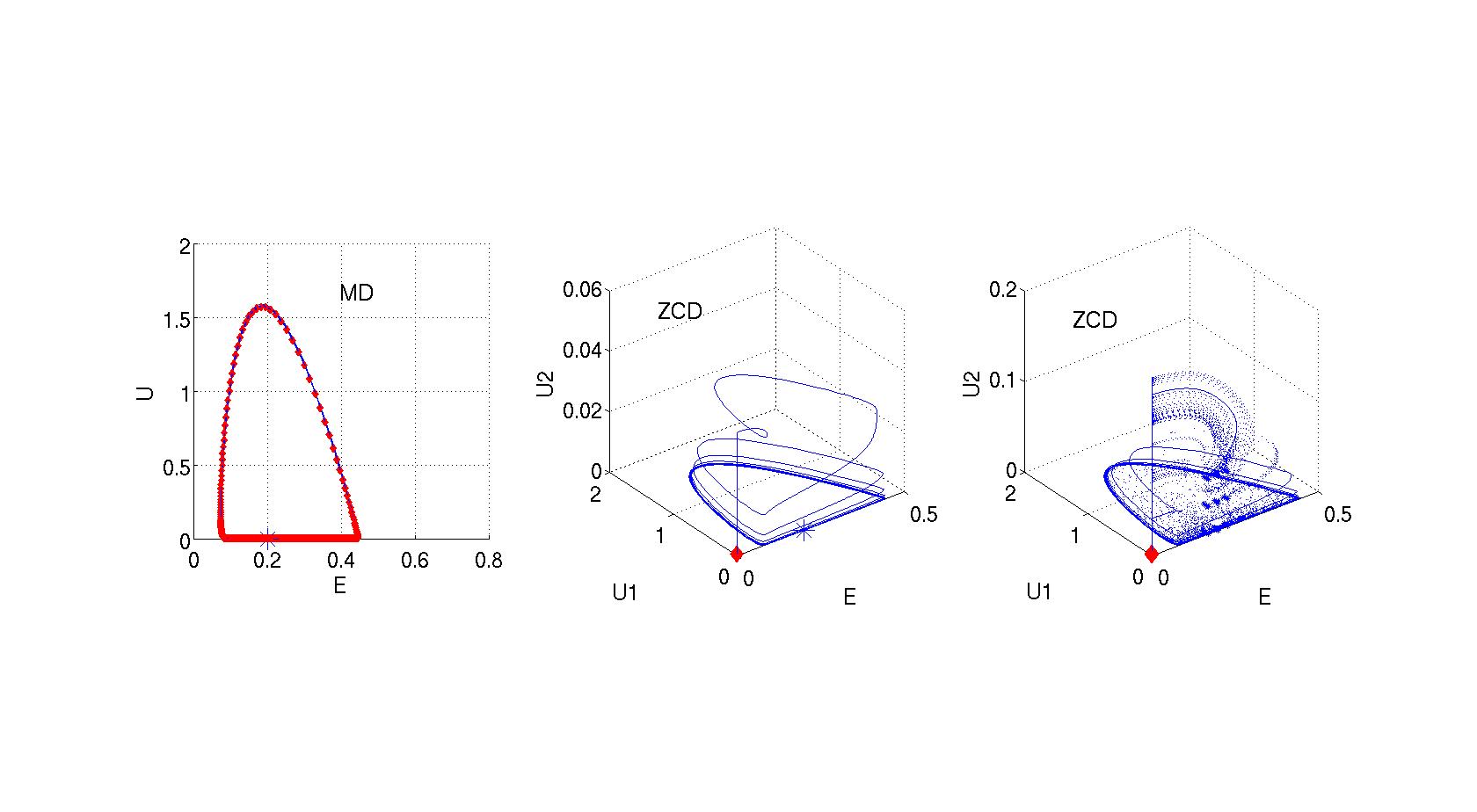}
\caption{First panel: Projection of Fig.3 in MD on $E$-$U$ plane. The parameters are $\nu=19$, $\eta=0.12$, $q=0.58$, $\rho=0.55$, $\sigma=0.6$, $\zeta=1.7$. Middle panel: Phase plot of Fig.4. Last panel: Phase plot of Fig.4 with 81 initial conditions. Stars denote initial values, blue dots denote trajectories and red diamonds denote final states.}
\end{center}
\end{figure*}

\begin{figure*}
\begin{center}
\includegraphics[width=2\columnwidth]{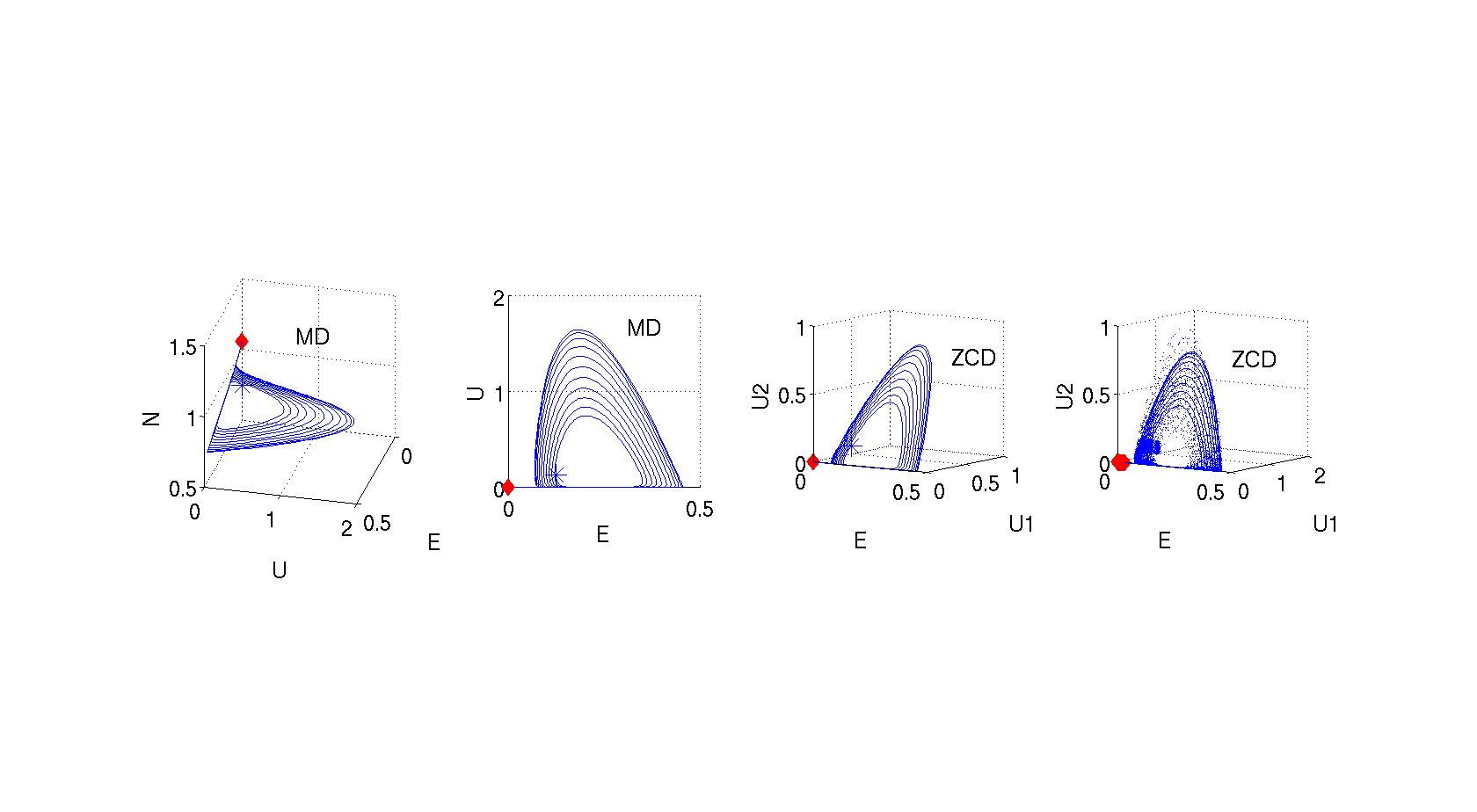}
\caption{First panel: Phase plot for Fig.4 of MD. Second panel: Projection of Fig.4 in MD on $E$-$U$ plane. The parameters are $\nu=19$, $\eta=0.12$, $q=0.582$, $\rho=0.55$, $\sigma=0.6$, $\zeta=1.7$. Third panel: Phase plot of Fig.5. Last panel: Phase plot of Fig.5 with 81 initial conditions. Stars denote initial values, blue dots denote trajectories and red diamonds denote final states.}
\end{center}
\end{figure*}

\begin{center}
\begin{table*}[ht]
{\small
\hfill{}
\begin{tabular}{|l|l|l|l|l|l|l|}
\hline
{Case}&{q}&{$\nu_{2}/\nu_{1}$}&{$\eta_{2}/\eta_{1}$}&{Timetraces}&{Phaseplot}&{Manifold} \\
\hline
1 & 0.47 & 1.01 & 1.01 & Fig.1 & Fig.7 & Fixed point\\ \cline{1-7}
2 & 0.47 & 0.01 & 0.1 & Fig.2 & Fig.8 & Limit cycle\\  \cline{1-7}
3 & 0.58 & 1.01 & 1.01 & Fig.3 & Fig.9 & Limit cycle\\  \cline{1-7}
4 & 0.58 & 0.01 & 0.01 & Fig.4 & Fig.10 & Limit cycle\\  \cline{1-7}
5 & 0.582 & 1.0001 & 1.0001 & Fig.5 & Fig.11 & Fixed point\\  \cline{1-7}
6 & 0.582 & 1.001 & 0.05;0.06;0.1;0.11 & Fig.6 & N/A & N/A\\
\hline
\end{tabular}}
\hfill{}
\caption{Summary of Figs.1 to 11}
\label{tb:tablename}
\end{table*}
\end{center}

\mbox{}\\
\textbf{APPENDIX: Identification and stability of fixed points}
\mbox{}\\

We start from Eqs.(5-8), and for simplicity define the normalized equations as

\begin{equation}
\begin{cases}
dE/dt = \left(N - N^4 - E - U_{1} - U_{2}\right) E \equiv f\left(E, U_{1}, U_{2}, N\right)\\
dU_{1}/dt  =  \nu_{1}\left( \dfrac{E}{1 + \zeta N^4} - \eta_{1}\right)U_{1} \equiv g_{1}\left(E, U_{1}, N\right)\\
dU_{2}/dt  = \nu_{2}\left( \dfrac{E}{1 + \zeta N^4} - \eta_{2}\right)U_{2} \equiv g_{2}\left(E, U_{2}, N\right)\\
dN/dt = q - \left( \rho + \sigma E\right)N \equiv h\left(E, N\right)
\end{cases}
\end{equation} 

We regard point $\left(N_{0}, E_{0}, U_{10}, U_{20} \right)$ as a fixed point of the 4D system, and define 

\begin{equation}
\begin{cases}
f_{0} \equiv  f\left(E_{0}, U_{10}, U_{20}, N_{0}\right) \\
g_{10} \equiv  g_{1}\left(E_{0}, U_{10}, N_{0}\right) \\
g_{20} \equiv  g_{2}\left(E_{0}, U_{20}, N_{0}\right) \\
h_{0} \equiv  h\left(E_{0}, N_{0}\right)
\end{cases}
\end{equation}

By construction $f_{0}=g_{10}=g_{20}=h_{0}=0$. Near the fixed point, we make a local linear expansion of the model parameters:

\begin{equation}
\bigtriangleup E \equiv E - E_{0};
\bigtriangleup U_{1} \equiv  U_{1} - U_{10};
\bigtriangleup U_{2} \equiv U_{2} - U_{20};
\bigtriangleup N \equiv  N - N_{0};
\end{equation}

This gives rise to the linearized equations

\begin{equation}
\begin{cases}
f \approx f_{0} + \frac{\partial f}{\partial E}\bigtriangleup E + \frac{\partial f}{\partial U_{1}}\bigtriangleup U_{1} + \frac{\partial f}{\partial U_{2}}\bigtriangleup U_{2} + \frac{\partial f}{\partial N}\bigtriangleup N \\
 g_{1} \approx g_{10} + \frac{\partial g_{1}}{\partial E}\bigtriangleup E + \frac{\partial g_{1}}{\partial U_{1}}\bigtriangleup U_{1} + \frac{\partial g_{1}}{\partial N}\bigtriangleup N \\
g_{2} \approx g_{20} + \frac{\partial g_{2}}{\partial E}\bigtriangleup E + \frac{\partial g_{2}}{\partial U_{2}}\bigtriangleup U_{2} + \frac{\partial g_{2}}{\partial N}\bigtriangleup N \\
h \approx h_{0} + \frac{\partial h}{\partial E}\bigtriangleup E + \frac{\partial h}{\partial N}\bigtriangleup N 
\end{cases}
\end{equation}
 
To obtain the eigenvalues of the system, we calculate the corresponding Jacobian matrix 

\begin{eqnarray}
J &=& \left( \begin{array}{rrrr} \frac{\partial}{\partial E}f & \frac{\partial}{\partial U_{1}}f & \frac{\partial}{\partial U_{2}}f & \frac{\partial}{\partial N}f \\ \frac{\partial }{\partial E}g_{1} & \frac{\partial }{\partial U_{1}}g_{1} & \frac{\partial }{\partial U_{2}}g_{1} & \frac{\partial }{\partial N}g_{1} \\ \frac{\partial }{\partial E}g_{2} & \frac{\partial }{\partial U_{1}}g_{2} & \frac{\partial }{\partial U_{2}}g_{2} & \frac{\partial }{\partial N}g_{2} \\ \frac{\partial}{\partial E}h & \frac{\partial}{\partial U_{1}}h & \frac{\partial}{\partial U_{2}}h & \frac{\partial}{\partial N}h  
\end{array} \right)_{\left(E_{0}, U_{10}, U_{20}, N_{0}\right)}
\end{eqnarray} 

We now identify the fixed points.

\textcircled{1} if $E = 0$,

\begin{equation}
\begin{cases}
N - N^4 - E - U_{1} - U_{2} = Anyvalue \\
U_{1} = 0 \\
U_{2} = 0 \\
N = \dfrac{q}{\rho}
\end{cases}
\end{equation}

\textcircled{2} if $E \neq 0$,

\begin{equation}
\begin{cases}
N - N^4 - E - U_{1} - U_{2} = 0 \\
\left( \dfrac{E}{1 + \zeta N^4} - \eta_{1}\right)U_{1} = 0 \\
\left( \dfrac{E}{1 + \zeta N^4} - \eta_{2}\right)U_{2} = 0 \\
q - \left( \rho + \sigma E\right)N = 0
\end{cases}
\end{equation}
 
From the second and third equations in this group, it follows that $U_{1}$ and $U_{2}$ cannot be non-zero simultaneously.

(i) if $U_{1} = 0$, $U_{2} \neq 0$, $E \neq 0$,

\begin{equation}
\begin{cases}
N - N^4 - E - U_{1} - U_{2} = 0 \\
\dfrac{E}{1 + \zeta N^4} - \eta_{1} = Anyvalue \\
\dfrac{E}{1 + \zeta N^4} - \eta_{2} = 0 \\
q - \left( \rho + \sigma E\right)N = 0
\end{cases}
\end{equation}

(ii) if $U_{1} \neq 0$, $U_{2} = 0$, $E \neq 0$,

\begin{equation}
\begin{cases}
N - N^4 - E - U_{1} - U_{2} = 0 \\
\dfrac{E}{1 + \zeta N^4} - \eta_{1} = 0 \\
\dfrac{E}{1 + \zeta N^4} - \eta_{2} = Anyvalue \\
q - \left( \rho + \sigma E\right)N = 0
\end{cases}
\end{equation}
 
(iii) if $U_{1}=U_{2}=0$, $E \neq 0$,

\begin{equation}
\begin{cases}
N - N^4 - E = 0 \\
U_{1} = 0 \\
U_{2} = 0 \\
q - \left( \rho + \sigma E\right)N = 0
\end{cases}
\end{equation}

Solutions for the specific cases of the MD and ZCD systems considered in this paper are shown in Tables II and Table III.

\newpage

\begin{figure}
\begin{center}
\includegraphics[width=1\columnwidth]{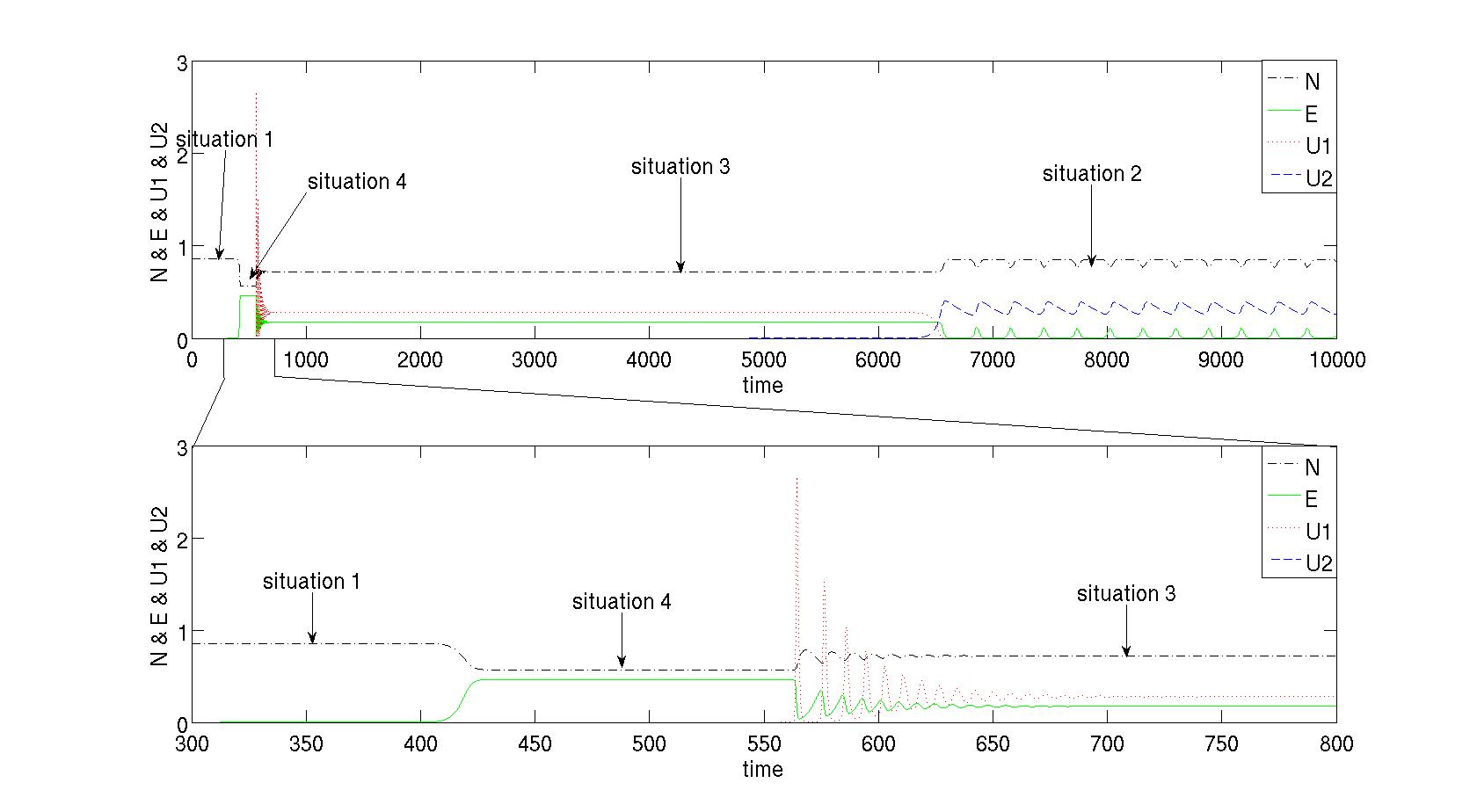}
\caption{Time series of Fig. 2 in this paper, annotated in light of Fig.13.}
\end{center}
\end{figure}

\begin{figure}
\begin{center}
\includegraphics[width=1\columnwidth]{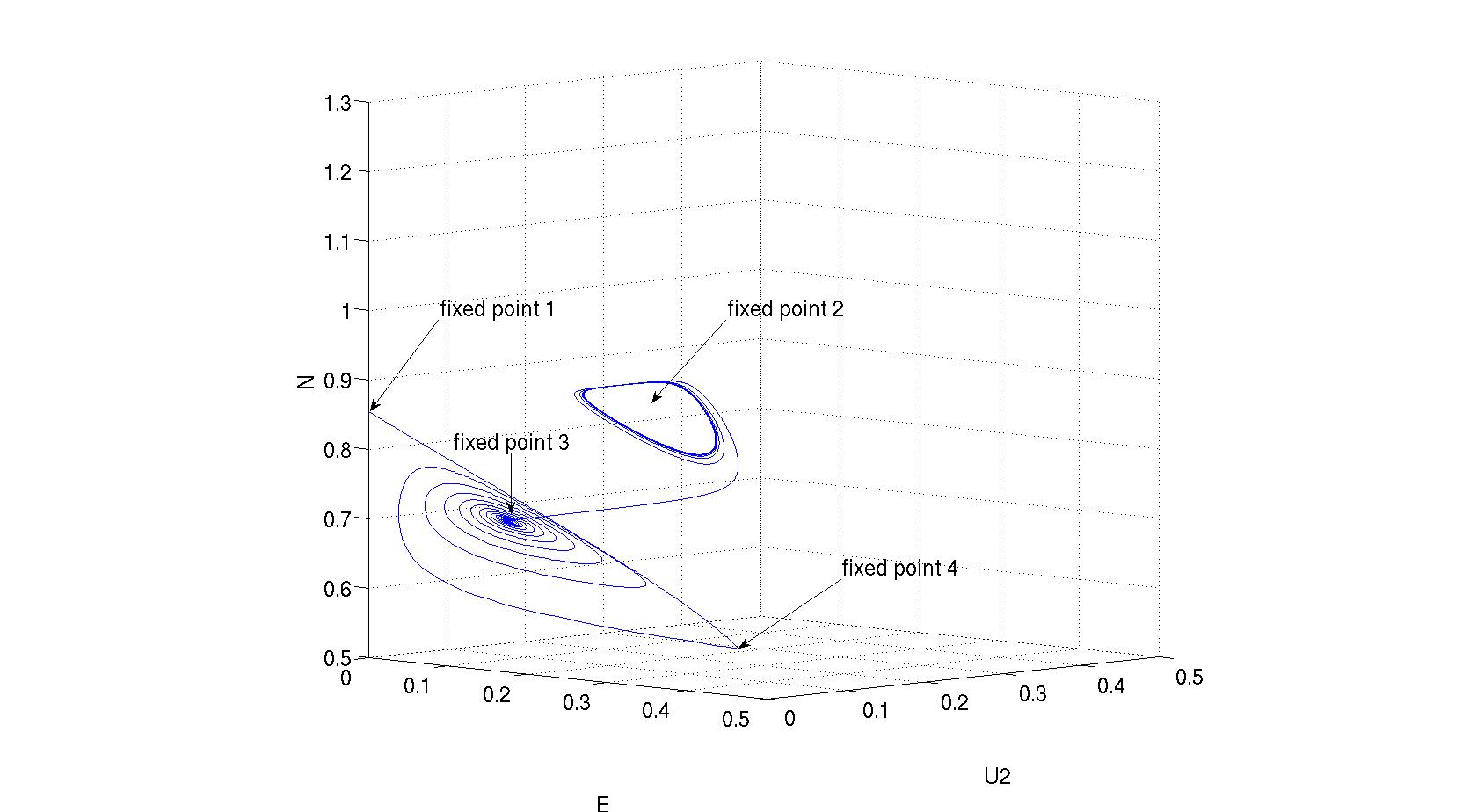}
\caption{Phase plot of Fig. 2 in this paper.}
\end{center}
\end{figure}

\begin{center}
\begin{table*}[ht]
{\small
\hfill{}
\begin{tabular}{|l|l|l|l|}
\hline
MD&Fixed points&Eigenvalues&Property \\
\hline
\multirow{3}*{Fig.2}
& $E=0$;$U=0$;$N=0.8545$ & -2.28;-0.55;0.3213 & Saddle point-Index 1  \\ \cline{2-4}
& $E=0.1742$;$U=0.2780$;$N=0.7181$ & $-0.0360\pm 0.8099i$;-0.7567 & Spiral node\bf{(final state)}  \\ \cline{2-4}
& $E=0.4638$;$U=0$;$N=0.5675$ & $-0.6460\pm 0.0963i$;5.2111 & Inward spiral and source\\ \cline{2-4}
\hline
\multirow{4}*{Fig.3}
& $E=0$;$U=0$;$N=0.8545$ & -2.28;-0.55;-0.1821 & Node  \\ \cline{2-4}
& $E=0.2249$;$U=0.1077$;$N=0.8468$ & $0.0069\pm 0.4991i$;$-0.9236$ & Outward spiral and sink\bf{(limit cycle)}  \\ \cline{2-4}
& $E=0.0769$;$U=0$;$N=0.9729$ & 0.0969;-0.7700;-1.7010 & Saddle point-Index 1 \\ \cline{2-4}
& $E=0.4588$;$U=0$;$N=0.7028$ & 3.8817;-0.3122;-0.9718 & Saddle point-Index 1 \\ \cline{2-4}
\hline
\multirow{4}*{Fig.4}
& $E=0$;$U=0$;$N=1.0582$ & -2.28;-0.55;-0.1957 & Node\bf{(final state)}  \\ \cline{2-4}
& $E=0.2260$;$U=0.1036$;$N=0.8489$ & $0.0080\pm 0.4892i$;-0.9275 & Outward spiral and sink  \\ \cline{2-4}
& $E=0.0825$;$U=0$;$N=0.9708$ & 0.1002;-0.7821;-1.6558 & Saddle point-Index 1   \\ \cline{2-4}
& $E=0.4576$;$U=0$;$N=0.7058$ & 3.8348;-0.3058;-0.9764 & Saddle point-Index 1  \\ \cline{2-4}
\hline
\end{tabular}}
\hfill{}
\caption{MD system}
\label{tb:tablename}
\end{table*}
\end{center}

\begin{center}
\begin{table*}[ht]
{\small
\hfill{}
\begin{tabular}{|l|l|l|l|}
\hline
ZCD&Fixed points&Eigenvalues&Property \\
\hline
\multirow{4}*{Fig.7}
& $E=0$;$U_{1}=0$;$U_{2}=0$;$N=0.8545$ & -2.28;-0.55;0.3213;-2.3258 & 4D Saddle point-Index 1  \\ \cline{2-4}
& $E=0.1757$;$U_{1}=0$;$U_{2}=0.2770$;$N=0.7171$ & $-0.0365\pm 0.8165i$;-0.7581;0.0228 & Inward spiral, source and sink  \\ \cline{2-4}
& $E=0.1742$;$U_{1}=0.2780$;$U_{2}=0$;$N=0.7187$ & $-0.0360\pm 0.8099i$;-0.0230;-0.7567 & Spiral node\bf{(final state)}\\ \cline{2-4}
& $E=0.4638$;$U_{1}=0$;$U_{2}=0$;$N=0.5675$ & $-0.6460\pm 0.0963i$;5.2402;5.2111 & Inward spiral and sources  \\ \cline{2-4}
\hline
\multirow{4}*{Fig.8}
& $E=0$;$U_{1}=0$;$U_{2}=0$;$N=0.8545$ & -2.28;-0.55;0.3213;-0.0023 & 4D Saddle point-Index 1  \\ \cline{2-4}
& $E=0.0219$;$U_{1}=0$;$U_{2}=0.3275$;$N=0.8346$ & $0.0019\pm 0.0272i$;-0.5888;-2.052 & Outward spiral and sinks\bf{(limit cycle)}  \\ \cline{2-4}
& $E=0.1742$;$U_{1}=0.2780$;$U_{2}=0$;$N=0.7187$ & $-0.0360\pm 0.8099i$;0.0205;-0.7567 & Inward spiral,source and sink \\ \cline{2-4}
& $E=0.4638$;$U_{1}=0$;$U_{2}=0$;$N=0.5675$ & $-0.6460\pm 0.0963i$;0.0726;5.2111 & Outward spiral and sources  \\ \cline{2-4}
\hline
\multirow{5}*{Fig.9}
& $E=0$;$U_{1}=0$;$U_{2}=0$;$N=1.0545$ & -2.28;-0.55;-2.3258;-0.1821 & Node \\ \cline{2-4}
& $E=0.2265$;$U_{1}=0$;$U_{2}=0.1078$;$N=0.8456$ & $0.0062\pm 0.5045i$;0.0228;-0.9248 & Outward spiral, source and sink  \\ \cline{2-4}
& $E=0.2249$;$U_{1}=0.1077$;$U_{2}=0$;$N=0.8468$ & $0.0069\pm 0.4991i$;-0.0230;-0.9236 & Outward spiral and sinks\bf{(limit cycle)} \\ \cline{2-4}
& $E=0.0769$;$U_{1}=0$;$U_{2}=0$;$N=0.9729$ & 0.0969;-0.7700;-1.7010;-1.741 & 4D Saddle point-Index 1  \\ \cline{2-4}
& $E=0.4588$;$U_{1}=0$;$U_{2}=0$;$N=0.7028$ & -0.3122;-0.9718;3.8817;3.8975 & 4D Saddle point-Index 2  \\ \cline{2-4}
\hline
\multirow{4}*{Fig.10}
& $E=0$;$U_{1}=0$;$U_{2}=0$;$N=1.0545$ & -2.28;-0.55;-0.1821;-0.0002 & Node\bf{(final state)}  \\ \cline{2-4}
& $E=0.2249$;$U_{1}=0.1077$;$U_{2}=0$;$N=0.8468$ & $0.0069\pm 0.4990i$;0.0226;-0.9236 & Outward spiral, source and sink \\ \cline{2-4}
& $E=0.4588$;$U_{1}=0$;$U_{2}=0$;$N=0.7028$ & -0.3122;-0.9718;3.8817;0.0614 & 4D Saddle point-Index 2  \\ \cline{2-4}
& $E=0.0769$;$U_{1}=0$;$U_{2}=0$;$N=0.9729$ & 0.0969;0.0056;-0.7700;-1.7010 & 4D Saddle point-Index 2  \\ \cline{2-4}
\hline
\multirow{5}*{Fig.11}
& $E=0$;$U_{1}=0$;$U_{2}=0$;$N=1.0582$ & -2.28;-0.55;-2.2805;-0.1957 & Node\bf{(final state)}  \\ \cline{2-4}
& $E=0.2260$;$U_{1}=0$;$U_{2}=0.1036$;$N=0.8489$ & $0.0080\pm 0.4892i$;0.0002;-0.9275 & Outward spiral, source and sink \\ \cline{2-4}
& $E=0.2260$;$U_{1}=0.1036$;$U_{2}=0$;$N=0.8489$ & $0.0080\pm 0.4892i$;-0.0002;-0.9275 & Outward spiral and sinks  \\ \cline{2-4}
& $E=0.0825$;$U_{1}=0$;$U_{2}=0$;$N=0.9708$ & 0.1002;-0.7821;-1.6558;-1.6562 & 4D Saddle point-Index 1  \\ \cline{2-4}
& $E=0.4576$;$U_{1}=0$;$U_{2}=0$;$N=0.7058$ & -0.3058;-0.9764;3.8348;3.8349 & 4D Saddle point-Index 2  \\ \cline{2-4}
\hline
\end{tabular}}
\hfill{}
\caption{ZCD system}
\label{tb:tablename}
\end{table*}
\end{center}

\end{document}